\documentstyle[12pt]{article}
\def\beq{\begin{equation}}
\def\eeq{\end{equation}}
\def\bea{\begin{eqnarray}}
\def\eea{\end{eqnarray}}
\def\bq{\begin{quote}}
\def\eq{\end{quote}}

\def\ATP{{\it Astropart. Phys.} }

\def\CQG{{\it Class. Quantum Gravity} }

\def\IJMP{{\it Int. J. Mod. Phys.} }

\def\MPL{{\it Mod. Phys. Lett.} }

\def\NP{{\it Nucl. Phys.} }
\def\PL{{\it Phys. Lett.} }
\def\PR{{\it Phys. Rev.} }
\def\PRL{{\it Phys. Rev. Lett.} }

\def\RMP{{\it Rev. Mod. Phys.} }

\def\gappeq{\mathrel{\rlap {\raise.5ex\hbox{$>$}}
{\lower.5ex\hbox{$\sim$}}}}

\def\lappeq{\mathrel{\rlap{\raise.5ex\hbox{$<$}}
{\lower.5ex\hbox{$\sim$}}}}

\begin{document}
\pagestyle{empty}

\begin{flushright}
{\sc DF/IST-1.98} \\
{\sc MIT-CTP$\sharp$2786} \\
{\sc October 1998} \\
{\tt gr-qc/9810013}
\end{flushright}

\input epsf

\vspace*{1.0cm}

\begin{center}
{\large\bf MODULAR QUANTUM COSMOLOGY}\\
\vspace*{1.0cm}
{\bf Orfeu Bertolami}\footnote{E-mail: {\tt orfeu@cosmos.ist.utl.pt}}\\
\medskip
{Departamento de F\'\i sica}\\
{Instituto Superior T\'ecnico, Av. Rovisco Pais}\\
{1096 Lisboa Codex, Portugal}\\
\vspace*{1.0cm}
and \\
\vspace*{1.0cm}
{\bf Ricardo Schiappa}\footnote{E-mail: {\tt ricardos@ctp.mit.edu}}\\
\medskip
{Center for Theoretical Physics and Department of Physics}\\
{Massachusetts Institute of Technology, 77 Massachusetts Ave.}\\
{Cambridge, MA 02139, U.S.A.}\\ 

\vspace*{1.5cm}
{\bf ABSTRACT} \\ 
\end{center}
\indent

\setlength{\baselineskip}{0.7cm}

We study solutions of the Wheeler-DeWitt equation corresponding to an 
$S$-modular invariant ${\cal N}=1$ supergravity model and a closed 
homogeneous and isotropic Friedmann-Robertson-Walker spacetime. The 
issues of inflation and the cosmological constant problem are 
addressed with the help of the relevant wave functions.
We find that topological type inflation is consistent from the 
quantum mechanical point of view and that a solution for the
cosmological constant problem along the lines of the strong CP problem
naturally arises.

\vfill
\eject

\setcounter{page}{2}
\pagestyle{plain}

\vspace{0.5cm}

\section{Introduction}

\indent

Duality symmetries play a fundamental role in modern string theories.
These symmetries made it possible to understand that all five known
string theories, as well as eleven-dimensional supergravity, are the weakly
coupled limit of a single and more comprehensive structure named $M$-Theory.
The basic dualities are of two types and are referred to as $T$-dualities and
$S$-dualities. Moreover, these dualities are combined in a more general 
symmetry, called $U$-duality. These transformations relate different, however 
equivalent, string theories \cite{Schwarz}. The first example encountered 
of such a symmetry, named target space modular invariance or $T$-duality, 
was the $O(d,d)$ transformation connecting all toroidal compactifications in 
$d$-dimensions \cite{Narain}. In this case, it was shown that the duality 
symmetry holds to all orders in the string loop expansion parameter, through 
a suitable change in the dilaton when transforming both metric and torsion 
fields \cite{Alvarez}. Furthermore, it was shown that the effective 
supergravity action following from string compactifications on orbifolds or 
Calabi-Yau manifolds is constrained by an underlying string symmetry, the 
mentioned target space modular invariance. The target space modular group 
$PSL(2,{\bf Z})$ acts on the complex scalar field T as 

\begin{equation}
T \to \frac{a T - i b}{i c T + d}; \qquad a, b, c, d \in {\bf Z}, 
\quad ad-bc=1,
\label{eq:1.1}
\end{equation}

\noindent
where $\langle T\rangle$ is the background modulus associated to the overall 
scale of the internal six-dimensional space on which the string theory is 
compactified, $T=R^2 + i B$, with $R$ being the ``radius'' of the internal 
space and $B$ an internal axion. The target space modular transformation 
contains the well-known duality transformation $R\to 1/R$, as well as discrete 
shifts of the axionic background $B$, and it was shown that this symmetry 
remains unbroken at any order in string perturbation theory. 

An important subset of these duality symmetries is the so-called scale-factor 
or Abelian duality  of string models embedded in flat homogeneous and 
isotropic spacetimes \cite{Veneziano}. The scale-factor duality symmetry
is present in the lowest order string effective action, implying that the
transformation of the scale factor of a homogeneous and isotropic
target space metric, $a(t) \to \pm a^{-1}(t)$, would leave the model invariant
provided that, in $d$ spatial dimensions, the string coupling -- the dilaton 
$\phi$ -- is transformed as

\begin{equation}
\phi(t) \rightarrow \phi(t) - {d\over 2} \ln a(t)~.
\label{eq:1.2}
\end{equation}

Other transformations were also proposed to implement these dualities for 
backgrounds with non-Abelian isometry groups which are, in principle, 
compatible with homogeneous Bianchi cosmological backgrounds \cite{Ossa}.

Already at the classical level, string theory allows for cosmological 
models with scale-factor duality symmetry from which important 
issues such as the problem of the initial singularity, inflation and the 
generation of primordial density fluctuations and gravitational waves
can be addressed. Scale-factor duality leads also to interesting 
cosmological scenarios as, for instance, the Pre-Big-Bang \cite{Gasperini} 
(see \cite{Veneziano1} and references therein for an updated account) which 
assumes the Universe has undergone a period of accelerated contraction 
towards the Big-Bang singularity and emerged due to yet unknown 
stringy effects in the expanding standard radiation dominated 
Friedmann-Robertson-Walker phase. This contracting phase has presumably
been driven by the dilaton kinetic energy and would give origin to a
substantial amount of gravitational waves which would then be a distinct 
observational signature of this scenario. Nevertheless, despite 
its appealing features, the Pre-Big-Bang scenario is plagued, 
at least in its simplest versions, with serious inconsistencies 
such as fine-tuning and instabilities that invalidate the tree level picture 
\cite{Hwang} and also with the lacking of a mechanism ensuring the transition 
from the Pre-Big-Bang phase towards the standard hot Big-Bang model 
\cite{Kaloper}. Actually, obtaining a period of inflation that emerges 
naturally from string theory is known to be a notoriously hard problem and 
many suggestions have been proposed \cite{Damour,Macorra,Bento2,Bento3}. 
We shall discuss in this paper a proposal in the context of dual 
${\cal N} = 1$ supergravity that is based on the idea of topological 
inflation and see how quantum cosmology does actually support some of its 
assumptions. Anyway, independently from the above mentioned difficulties, 
it is possible to show, for instance, that string cosmological models where
the scale-factor duality symmetry holds naturally allow for an evolution
towards a radiation-dominated phase \cite{Tseytlin}. Of course, a
consistent treatment of issues related with the very early Universe 
requires understanding of the higher curvature regime, where quantum gravity 
effects are important, and the question of whether string symmetries still 
hold in the quantized version of the theory is quite relevant. 
As a complete quantum field theory of closed strings is notoriously 
difficult to handle, one hopes to get some insight into the full theory by
considering the quantum cosmology of the low-energy effective action (for 
a discussion on the validity and consistency of the effective action procedure
in quantum cosmology see, for instance, Ref. \cite{Bertolami1}). The issue of 
whether scale-factor duality survives at the quantum level, was considered 
via the canonical quantization of the lowest order string effective action 
using the standard ADM formalism in an ${\bf R}\times S^3$ topology in 
Ref. \cite{Bento1}. There, it was shown within the formalism of quantum 
cosmology and its interpretative framework \cite{Hartle} that, although 
scale-factor duality is lost as an exact symmetry of the resulting 
minisuperspace model, it still holds as an approximate symmetry of the 
classical string model, as the wave function was shown to peak for field 
configurations consistent with this symmetry. The analysis of the one-loop 
string effective models which exhibit the full $O(d,d)$ symmetry was 
studied in Ref. \cite{Kehagias}.
Furthermore, the quantum treatment has also been 
considered to address the question of whether a quantum transition would 
allow for an exit from the Pre-Big-Bang phase to the standard radiation 
dominated phase \cite{Gasperini1}. We mention in relation to
these issues, that the conditions under which 
quantum cosmology based on the low-energy Einstein-Hilbert action arises
from a subgroup of the modular group of $M$-theory, as well as how duality 
transformations can resolve apparent cosmological singularities
has been recently discussed in Ref. \cite{Banks}. 

$S$-duality was conjectured \cite{Font} in analogy with $T$-duality. This 
conjectured symmetry would be a further modular invariance in the 
resulting ${\cal N}=1$ supergravity model arising from string theory, where 
the modular group acts on the complex scalar field (which is
the lowest order component of a chiral superfield in the 4-dimensional 
string), $S=\phi + i \chi $, where $\chi$ is a pseudoscalar (axion) field. 
This symmetry includes a duality invariance under which the dilaton gets 
inverted, the so-called $S$-duality, that is strong-weak coupling duality. 
$S$-modular invariance strongly constrains the theory since it relates the 
weak and strong coupling regimes as well as the ``$\chi$-sectors'' of the 
theory. This symmetry was also conjectured in ${\cal N}=4$ supersymmetric 
four-dimensional theories \cite{Sen}.

For further convenience, we outline here the most relevant features of the 
Hartle-Hawking proposal \cite{Hartle}. In quantum cosmology it is assumed 
that the quantum state of a 4-dimensional Universe is described by 
a wave function 
$\Psi[h_{ij}, \Phi_0]$, which is a functional of the spatial 3-metric, 
$h_{ij}$, and of the matter fields, generically denoted by $\Phi_0$, on a 
compact 3-dimensional hypersurface $\Sigma$. The hypersurface $\Sigma$ is then 
regarded as the boundary of a compact 4-manifold $M^4$ on which the 4-metric 
$g_{\mu\nu}$ and the matter fields are regular. The metric $g_{\mu\nu}$ and 
the fields $\Phi$ coincide with $h_{ij}$ and $\Phi_0$ on $\Sigma$ and the 
wave function is then defined through the path integral over 4-metrics, 
$^{4}g$, and matter fields:

\begin{equation}
\Psi[h_{ij}, \Phi_0] =
\int_{{\cal C}} {\cal D}[^{4}g] {\cal D}[\Phi] \exp\left(-S_{E}[^{4}g, 
\Phi]\right) ~, \label{eq:pathint}
\end{equation}

\noindent
where $S_E$ is the Euclidean action and $\cal C$ is the class of 4-metrics 
$g_{\mu\nu}$ and regular fields $\Phi$ defined on Euclidean compact 
manifolds $M^4$ and which have no boundary other than $\Sigma$.

We stress that since the quantum cosmology approach of Hartle and Hawking 
allows for a well defined programme for establishing this set of initial 
conditions, it is quite natural to consider it in studying unified 
supergravity models arising from string theory. This programme has been 
already applied to many different models of interest such as massive scalar 
fields \cite{Hawking1}, Yang-Mills fields \cite{Bertolami2}, massive vector 
fields \cite{Bertolami3} as well as in supersymmetric models (see 
Ref. \cite{Moniz} for a review and a complete set of references) and 
multidimensional Einstein-Yang-Mills theories with $SO(N)$ gauge groups 
\cite{Bertolami4}. In this work we shall use the quantum cosmology approach 
to further study and confirm the assumptions and conditions 
under which energy density fluctuations and gravitational waves were 
generated (or were modestly generated in the case of gravitational waves)
in the context of inflation of the topological type, as  
discussed in Refs. \cite{Bento2,Bento3}, within ${\cal N}=1$ 
supergravity models with $S$ and $T$ dualities \cite{Font,Macorra}. 
Indeed, it was argued in Refs. \cite{Bento2,Bento3} that domain walls 
separating inequivalent vacua could, in $S$-dual and in some
$S$ and $T$-dual ${\cal N}=1$ supergravity models, inflate 
and that in this process energy density fluctuations and gravitational 
waves could be generated -- provided that the relevant fields were close to 
the local maximum of the potential. We shall see that this hypothesis is 
actually confirmed. We shall also study the behavior of the wave function 
of the Universe in the large scale factor limit, in order to address the 
cosmological constant problem and confront it with the arguments put 
forward in Ref. \cite{Kar} where it was discussed the role played by 
$S$-duality in the vanishing of a bare tree level cosmological constant.
     
The organization of this paper is the following. In section 2 we introduce 
the relevant features of the modular invariant structures in ${\cal N}=1$ 
supergravity and of the closed homogeneous and isotropic 
Friedmann-Robertson-Walker spacetime which is going to be our stage for 
studying $S$-modular invariance. We subsequently set up the minisuperspace 
model of our analysis and after solving the classical constraints and the 
canonical conjugate momenta we obtain the Wheeler-DeWitt equation. In 
section 3 we study the boundary conditions of the  Wheeler-DeWitt equation 
and obtain solutions in the limit of small scale-factor, $a$, and large $a$. 
These solutions will allow us to discuss the issues of initial conditions for 
the $S$ field in topological inflation and the problem of the smallness of 
the cosmological constant. In section 4 we shall consider the interpretation 
of the wave function in the various regimes that have been studied in 
section 3. Finally, in section 5 we discuss our results and present our 
conclusions.

\vspace*{0.5cm}

\section{Effective Model and Wheeler-DeWitt Equation}

\indent

In this section we describe our minisuperspace model, which arises when
considering an $S$-modular invariant ${\cal N} = 1$ supergravity theory 
in a closed homogeneous and isotropic Friedmann-Robertson-Walker spacetime. 
The resulting model can be regarded as the one emerging in the field 
theory limit of heterotic string theory or the weak string coupling limit 
of $M$-Theory. Of course, it is arguable to consider $S$-modular invariance in 
${\cal N}=1$ supergravity theories as this invariance is shown to hold only 
in theories with more supersymmetries. However, given the importance of the 
$S$-modular invariance in string theory and of supergravity in all 
phenomenologically viable extensions of the Standard Model, our quantum 
analysis will refer all to an $S$-modular invariant ${\cal N} = 1$ 
supergravity theory. Moreover, as our main purpose is to gain insight into 
the dilaton-gravity physics we shall consider only the $NS\otimes NS$ bosonic 
part of the supergravity action. The bosonic action is given in terms of 
$S$ and $S^{+}$ fields and gravity, as \cite{Font}: 

\begin{equation}
S[g_{\mu \nu}, S, S^{+}] = 
\int_{M^4} d^4 {x} \sqrt{-g}~\left[~R 
+ {1 \over (S + S^{+})^{2}} 
\partial_{\mu} S~\partial^{\mu} S^{+} - V(S, S^{+})\right]~, 
\label{eq:2.1}
\end{equation}

\noindent
where $g$ is $\det \left(g_{\mu\nu}\right)$, $g_{\mu\nu}$ is the 4-dimensional 
metric, $R$ is the scalar curvature and we have set $\frac{M_{P}}{\sqrt{8 
\pi}} \equiv 1$. The potential $V(S, S^{+})$ is given in terms of 
$S$-invariant modular functions:

\begin{equation}
V(S,S^{+}) = \frac{1}{S_R\vert \eta(S)\vert^{4}}
\left( \frac{S_R^2}{4 \pi^2}\vert \hat G_2(S) \vert^2 - 3 \right),
\label{eq:2.1a}
\end{equation}

\noindent
where $S_R = 2~{\bf Re}~S$. The function $\eta(S)=q^{1/24} \prod_n(1-q^n)$ 
is the Dedekind function, $q\equiv \exp (-2\pi S)$; $\hat G_2(S) = G_2(S) - 
2 \pi/S_R$ is the weight two Eisenstein function and $G_2(S)=\frac{1}{3} 
\pi^2 - 8 \pi^2  \sum_{n} \sigma_1(n) \exp(- 2\pi n S)$, 
where $\sigma_1(n)$ is the sum of the divisors of $n$. The potential 
(\ref{eq:2.1a}) is shown in Figure 1.

To this potential (\ref{eq:2.1a}) one has to add the contribution of $D$-terms 
associated with the gauge sector of the theory. We shall assume that these 
fields are in their ground state and hence the contribution from $D$-terms 
will amount to a constant contribution to the potential (\ref{eq:2.1a}) once 
the field $S$ itself is settled in its ground state. The contribution from 
the $D$-terms can be written in an $S$-modular invariant form,

\begin{equation}
\label{new}
V_D = \frac{1}{2~{\bf Re}~f}\, D^2 \ ,
\label{eq:2.1b}
\end{equation}

\noindent
where $D=\hat g K^i {T_i}^j \Phi_j + \xi$, $\hat g$ being the gauge charge, 
${T_i}^j$ are the generators of the gauge group and $\xi$ is the 
Fayet-Illiopoulos term. $S$-modular invariance is ensured for 
$f = \frac{1}{2 \pi}[\ln(j(S) - 744]$, $j(S)$ being the 
generator of modular invariant functions and where for large $S$ one has
$j(S) = \frac{1}{q} + 744 + 196884 q + O(q^2)$, with $q = e^{-2 \pi q}$ 
\cite{Macorra,Lalak}. From string perturbative results it follows that
$f=S$ and hence, $S$-duality implies that $f\rightarrow f$. Another possible 
realization for ensuring $S$-duality is $f\rightarrow 1/f$, although this 
requires the existence of the so-called ``magnetic condensate'' 
\cite{Font,Lalak}.

Let us now turn to the discussion of the geometrical setting of our model. 
We shall restrict ourselves to spatially homogeneous and isotropic field 
configurations. A general discussion of the field configurations 
associated with the geometry we shall use, based on the theory of symmetric 
fields, can be found in Refs. \cite{Bertolami5,Bento4}. The most general form 
of the metric is 

\begin{equation}
ds^{2} = - N^2(t) dt^2 + a^2(t) 
\sum_{i=1}^3 \omega^i \omega^i~, 
\label{eq:2.2}
\end{equation}

\noindent
where the scale factors $a(t)$ and the lapse function $N(t)$ are arbitrary 
non-vanishing functions of time, $\omega^\alpha$ denote local moving 
coframes in $S^3$ and $\sum_{i=1}^3 \omega^i \omega^i$ coincides with the 
standard metric $d \Omega_3^2$ of a 3-dimensional sphere.

Consistency with the geometry requires that the field $S$ depends just on 
time:

\begin{equation}
S(t, x^i) = S (t).
\label{eq:2.3}
\end{equation}

Substituting the ans\"atze (\ref{eq:2.2}) and (\ref{eq:2.3}) into the action 
(\ref{eq:2.1}), we obtain a one-dimensional effective action:

\begin{equation}
S_{\rm eff} [a, S, S^{+}]  = 
2 \pi^2 \int dt~\left[6~a~N(\left(\frac{\dot{a}}{N}\right)^{2} - 1) - 
\frac{N~a^{3}}{(S + S^{+})^2}~\frac{\dot{S}}{N}~\frac{\dot{S^{+}}}{N} - 
N~a^{3}~\hat V(S,S^{+})\right]~,
\label{eq:2.4}
\end{equation}

\noindent
where $\hat V(S,S^{+}) = V(S,S^{+}) + V_D$ for constant gauge fields $\Phi_j$.

The canonical conjugate momenta associated with the canonical variables 
are the following:

\begin{equation}
\Pi_a  =  24 \pi^{2}~a~\frac{\dot{a}}{N}~,~
\Pi_{S} = - 2 \pi^2 \frac{a^3}{(S + S^{+})^2}~\frac{\dot{S^{+}}}{N}~,~
\Pi_{S^{+}} = - 2 \pi^2 \frac{a^3}{(S + S^{+})^2}~\frac{\dot{S}}{N}~.~
\label{eq:2.5} 
\end{equation}

The minisuperspace Hamiltonian density, in the $N= 1$ gauge, is given by:

\begin{eqnarray}
{\cal H} & = & \Pi_{a}~\dot{a} +  \Pi_{S}~\dot{S} +  \Pi_{S^{+}}~\dot{S^{+}} - 
{\cal L} \nonumber \\
\\ \nonumber
& = & \frac{1}{48 \pi^{2} a}~\Pi_{a}^2 - 
\frac{(S + S^{+})^2}{2 \pi^{2} a^{3}}~
\Pi_{S}~\Pi_{S^{+}} + 2 \pi^{2} a (6 + a^{2}~\hat V(S,S^{+}))~. 
\label{eq:2.6}
\end{eqnarray} 

Canonical quantization then follows by promoting the conjugate momenta 
into operators: 

\begin{equation}
\Pi_a \mapsto -i\frac{\partial}{\partial \, a}~,
\Pi_S \mapsto -i\frac{\partial}{\partial \, S}~,
\Pi_{S^{+}} \mapsto -i\frac{\partial}{\partial \, S^{+}}~.
\label{eq:2.7}
\end{equation}

Finally, one obtains the Wheeler-DeWitt equation: 

\begin{equation}
\left[\frac{\partial^2}{\partial a^2} -
\frac{24~(S + S^{+})^2}{a^2} \frac{\partial^2}{\partial S~\partial S^{+}} -  
96~\pi^{4}~a^{2}~(6 + a^{2}~\hat V(S,S^{+}))\right] \Psi[a,S,S^{+}] = 0~,
\label{eq:2.8}
\end{equation}

\noindent
where in the usual parameterization of the factor ordering ambiguity, 
$\pi_a^2 \mapsto - a^{-p} \frac{\partial}{\partial a} \left(a^p  
\frac{\partial}{\partial a} \right)$, we have set $p=0$. Of course, our 
results will not depend on this choice as a change in the ordering of 
operators amounts only to a change in the normalization of the wave function.

Given the importance of the $S$-modular invariance in the ${\cal N}=1$ 
supergravity model we are studying, it would be more than natural to consider 
the wave function of the Universe in terms of explicitly $S$-invariant 
modular forms. Actually, the conservation law associated to $S$-modular 
invariance seems to imply that $S$ and $S^{+}$ should appear in the wave 
function only through $S$-modular invariant combinations. The $S$-modular 
invariant forms present in our model via the potential (\ref{eq:2.1a}), 
are the following:

\begin{equation}
X(S,S^{+}) \equiv (S + S^{+}) \vert \eta(S) \vert^{4} = 
S_R \vert \eta(S) \vert^{4}~
\label{eq:2.10}
\end{equation}

\noindent 
and

\begin{equation}
Y(S,S^{+}) \equiv (S + S^{+})^{2} \vert \hat G_2(S) \vert^2 = 
S_R^{2}~\vert \hat G_2(S) \vert^2~.
\label{eq:2.11}
\end{equation}

Hence, one might assume that $\Psi[a,S,S^{+}]=\Psi[a,{\cal X},{\cal Y}]$,
where ${\cal X} = {\cal X}(X,Y)$ and ${\cal Y} = {\cal Y}(X,Y)$ are
$S$-modular invariant combinations of $X$ and $Y$. This 
statement would imply, as shown in \cite{Kar}, that any bare cosmological 
constant would have to vanish in order to preserve the $S$-modular invariance. 
We shall see, however, that this is not so if the cosmological constant 
arises, as in our model, from an explicitly $S$-modular invariant potential 
and if in this process $S$-modular invariance is spontaneously broken down 
to some smaller symmetry -- as we shall discuss below. Before pursuing this 
issue, let us show that the dependence of the wave function on $S$-modular 
invariant structures is more involved than what we have previously 
suggested. Indeed, starting from the relation between $\eta(S)$ and 
$G_{2}(S)$ \cite{Font},

\begin{equation}
G_{2}(S) = - \frac{4 \pi}{\eta(S)}~ \frac{\partial \eta(S)}{\partial S}~, 
\label{eq:2.12}
\end{equation}

\noindent
one finds,

\begin{equation}
\frac{\partial^{2} X(S,S^{+})}{\partial S~\partial S^{+} } = 
\frac{1}{4 \pi^2}~X(S,S^{+})~\vert \hat G_2(S) \vert^2 ~, 
\label{eq:2.13}
\end{equation}

\noindent
that involves $G_2(S)$. This means that if we initially assume that the wave 
function of the Universe depends upon $X(S,S^{+})$, then it must also depend 
on $Y(S,S^{+})$. To realize that such is true, one just has to assume 
otherwise, and after a simple calculation such as 
 
\begin{equation}
\frac{\partial^{2} \Psi[a,X(S,S^{+})]}{\partial S~\partial S^{+}} 
= \frac{\vert \hat G_2(S) \vert^{2}}{4 \pi^2} \left[X^{2}~ 
\frac{\partial^{2} \Psi[a,X]}{\partial^{2} X} + 
X~\frac{\partial \Psi[a,X]}{\partial X}\right]~, 
\label{eq:2.14}
\end{equation}

\noindent
one immediately sees that this is not coherent as the Wheeler-DeWitt equation 
now involves more modular structures than initially assumed. So, our initial 
assumption of $\Psi[a,S,S^{+}]=\Psi[a,{\cal X},{\cal Y}]$, with 
${\cal X} = {\cal X}(X,Y)$ and ${\cal Y} = {\cal Y}(X,Y)$ seems reasonable. 
However, from a similar line of thought as the previous one, we shall see 
that even more modular structures are needed.

Indeed, as one computes the differential operators of the Wheeler-DeWitt 
equation under the previous assumption for the wave function, one finds 
that the derivative ${\partial \over \partial S}Y(S,S^{+})$ appears as part 
of such operators, and such derivative depends upon ${\partial \over \partial 
S}\hat{G}_2(S)$. Now, from the the covariant derivative of a modular form 
of weight $d$, $F_{d}$,  $D_{d} F_{d} = F_{d + 2}$, where $D_{d} \equiv 
\frac{i}{\pi} \frac{\partial}{\partial S} + \frac{d}{2 \pi S_{R}}$ it follows 
that 

\begin{eqnarray}
D_{2} \hat G_{2} = \frac{1}{6} G_{4} - \frac{1}{6}{\hat G_{2}}^2~~,~~~~~
D_{4} G_{4} = \frac{2}{3} G_{6} - \frac{2}{3} \hat G_{2} G_{4}~~,~~~~~
D_{6} G_{6} = G_{4}^2 - \hat G_{2} G_{6}~~,~
\label{eq:2.15}
\end{eqnarray} 

\noindent
and so on. This then implies that the derivatives of the modular structures 
according to $S$ and $S^{+}$, present in the Wheeler-DeWitt equation, will 
give origin to terms involving a new modular invariant form involving $G_{4}$, 
say $Z(S,S^{+}) \equiv (S + S^{+})^{4} \vert \hat G_4(S) \vert^2$. 
This should have to be considered in the wave function, implying that 
${\cal X} = {\cal X}(X,Y,Z)$ and ${\cal Y} = {\cal Y}(X,Y,Z)$. Of course, 
this new modular invariant structure via derivative terms would give origin 
to a modular structure involving $G_{6}$ which should be included into the 
wave function and so on. One concludes then that the whole Eisenstein series 
should be involved and that considering modular invariant structures to 
obtain the wave function of the Universe is not a very practical procedure.

In the next section we shall study the boundary conditions of the 
Wheeler-DeWitt equation and look for solutions of equation (\ref{eq:2.8})
that depend explicitly on $S$ and $S^{+}$ in the limits of small and large 
scale-factor. 

\vspace*{0.5cm}

\section{Solutions of the Wheeler-DeWitt Equation}

\indent

We start this section by establishing the boundary conditions for the 
Wheeler-DeWitt equation (\ref{eq:2.8}). We shall obtain these boundary 
conditions through the path integral representation for the ground-state of 
the Universe in a compact manifold $C$ \cite{Hartle},

\beq
\Psi[a,S,S^{+}]=\int_{C}{{\cal D}[a] {\cal D}[S] {\cal D}[S^{+}] 
\exp(-S_{\rm E})}~,
\label{eq:3.1}
\eeq

\noindent
which allows evaluating $\Psi(a,S,S^{+})$ close to $a = 0$. Notice that
since one can have, via an appropriate choice of the metric, 
$\Psi[a,S,S^{+}] = e^{-S_E}$ near the past null infinity $\cal I^{-}$, 
the procedure we are using here is really the most suitable one. The Euclidean 
action, $S_{\rm E}=-iS_{\rm eff}$, is obtained through the effective action 
(\ref{eq:2.4}) taking $d\tau=iNdt$  such that the Euclidean metric is compact 

\beq
\hat{ds^2}=d\tau^2+ a^{2}(\tau)\sum_{i=1}^3 \omega^i \omega^i~. 
\label{eq:3.2}
\eeq

In order to estimate $\Psi[a,S,S^{+}]$ close to $a \rightarrow 0$ one 
evaluates $S_E$ from $\tau = 0$ to $\Delta \tau$:

\beq
S_E= 2~\pi^{2}\int_{0}^{\Delta \tau} {d\tau \left[6~a~(\dot{a}^{2} + 1) - 
\frac{a^{3}}{S_R^{2}}~\dot{S}~\dot{S^{+}} + a^{3}~\hat V(S,S^{+}) \right]}~.
\label{eq:3.3}
\eeq 

Close to $\tau = 0$, $a(\tau) \approx \tau$, then

\beq
S_E= 2~\pi^{2}\int_{0}^{\Delta \tau} {d\tau \left[12~\tau - 
\frac{\tau^{3}}{S_R^{2}}~\dot{S}~\dot{S^{+}} + \tau^{3}~
\hat V(S,S^{+}) \right]}~,
\label{eq:3.4}
\eeq 

\noindent
which yields for regular $S$, $S^{+}$ and $\hat V(S,S^{+})$ and 
non-vanishing $S$ that $S_E \rightarrow 0$ as  $\Delta \tau \rightarrow 0$ 
and hence that $\Psi[a,S,S^{+}] \rightarrow 1$. For vanishing $S$ and for 
$S \to \infty$ it follows that $\hat V(S,S^{+}) \to \infty$ implying that 
$\Psi[a,S,S^{+}] \rightarrow 0$. 

In order to proceed one should also have to establish the regions where the
solution of the Wheeler-DeWitt equation is oscillatory or exponential. This 
can be done studying the regions where, for surfaces of constant 
minisuperspace potential $U \equiv 96\pi^{4}a^{2}(6 + a^{2}~\hat V(S,S^{+}))$, 
the minisuperspace metric $ds^2 = -~da^2 + dS~dS^{+}$ is either spacelike 
$ds^2 < 0$ or timelike $ds^2 > 0$ (see for instance \cite{Bertolami4} 
and references therein). However, as we are going to discuss soon, we shall 
use the scale-factor duality to obtain the very early Universe wave function 
from the very late Universe wave function and hence this study is not so 
crucial. Nevertheless, we shall discuss in section 4 how, via the study of 
the square of the trace of the extrinsic curvature, $K^2=K_{ij}K^{ij}$, one 
can determine whether the wave function corresponds to a Lorentzian or to an 
Euclidean geometry.

We are now ready to start studying solutions to the Wheeler-DeWitt equation. 
As an exact solution to the full equation is not possible to obtain, we 
shall study particular solutions under certain particular regimes of the 
evolution of the Universe. Namely we shall look at four distinct 
approximations: what we call the very early Universe, {\it i.e.}, when we have 
$a<<1$, and whose solution shall be denoted by $\Psi^{(--)}$; the early 
Universe, {\it i.e.}, when we have $a<1$ but not $a<<1$ 
(such that $a^2>>a^4$), and whose solution 
shall be denoted by $\Psi^{(-)}$; the late Universe, {\it i.e.}, when we have 
$a>1$, and whose solution shall be denoted by $\Psi^{(+)}$; and the very late 
Universe, {\it i.e.}, when we have $a>>1$ and $a^4>>a^2$, and whose solution 
shall be denoted by $\Psi^{(++)}$.

From the boundary condition one can search for solutions relevant for the
early Universe, namely when $a < 1$, but with $a^{2} >> a^{4}$ in
order to avoid the $a = 0$ curvature singularity. The wave 
function can be separated as $\Psi(a,S,S^{+}) = A(a)~F(S,S^{+})$ 
and a solution for the Wheeler-DeWitt equation is obtained in terms of Bessel
functions:

\beq
\Psi^{(-)}(a,S,S^{+}) =a^{1/2}\left\{k_{1}~
I_{-1/4}[12 \pi^{2} a^{2}] + k_{2}~
I_{1/4}[12 \pi^{2} a^{2}]\right\}~~, 
\label{eq:3.5}
\eeq 

\noindent
where the integration constants $k_{1}$ and $k_{2}$ are given in terms of
$~F(S,S^{+})$ which was maintained fixed. This implies that under the latter
conditions, that is fixed $~F(S,S^{+})$, the wave function predicts an
expanding Universe.

A solution for the very early Universe can be obtained from a
solution for the very late Universe through the scalar-factor duality,
with an appropriate shift in the real part of the $S$-field such that it 
corresponds to a sub-set of modular transformations (\ref{eq:1.1}) for this 
field. We obtain in this case the following wave function:  

\beq
\Psi^{(--)}(a,S,S^{+})  =  a^{-1/2}
\left\{c_{1}~I_{-1/6}[\frac{1}{3}(96\pi^{4}\hat V)^{1/2}a^{-3}] 
 +  c_{2}~I_{1/6}[\frac{1}{3}(96\pi^{4}\hat V)^{1/2}a^{-3}]\right\}~,
\label{eq:3.6}
\eeq 

\noindent
where $c_{1}$ and $c_{2}$ are constants.
This wave function indicates that the most likely configuration for the fields 
$S$ and $S^{+}$ is the one where they sit at the top of $\hat V$, as the 
probability of a given configuration is given by $\vert \Psi \vert ^{2}$.
This result is consistent with the assumption of Ref. \cite{Bento3} 
when computing the scalar and tensor perturbations, that is energy density 
fluctuations and gravitational waves, generated by the $S$ field during 
inflation. As discussed in \cite{Bento2}, inflation of the topological type 
can take place in models with $S$-modular invariance due to the presence of 
domain walls separating different vacua of the theory. This possibility has 
been discussed in generical terms by Linde \cite{Linde} and Vilenkin 
\cite{Vilenkin} and it was suggested in the context of string cosmology 
\cite{Banks1} in order to solve the Polonyi problem \cite{Coughlan} due to 
moduli fields. Thus, quantum cosmology does support the assumption considered 
in the topological inflationary model of \cite{Bento1,Bento2}, built in the 
context of an $S$-modular invariant ${\cal N} = 1$ supergravity, that the 
field $S$ starts at the top of $\hat V$ before inflation takes place.  
It should be pointed out that, as shown in Refs. \cite{Bento2,Bento3}, a 
realistic model requires that both $S$ and $T$ dualities are considered 
in order to match the amplitude of energy density fluctuations as observed 
by COBE. Moreover, it was shown in \cite{Bento3} that from the latter 
requirement and from COBE bounds for the spectral index of scalar 
perturbations, $0.7 < n_{s} < 1.2$, it follows that the spectrum of tensor 
perturbations is nearly flat, which is consistent with observations and that 
its amplitude is fairly modest $A_{t} \sim 10^{-2} A_{s}$. This contrasts, 
for instance, with what is expected from the Pre-Big-Bang model
\cite{Veneziano1}.

Let us now turn to the problem of the cosmological constant. It has been 
argued by many authors that $S$-duality may play an important role in the
vanishing or smallness of the cosmological constant. Indeed, for instance, 
Witten \cite{Witten1} has recently conjectured that via a duality 
transformation, a 4-dimensional field theory could be related to a 
3-dimensional one with the advantage that in the latter the breaking of 
supersymmetry, a condition imposed upon by phenomenology, does not imply that  
the cosmological constant is non-vanishing. In Ref. \cite{Kar}, it was shown 
that introducing a bare cosmological constant in the action implies that the 
resulting equations of motion lose the invariance under $S$-duality, unless 
one has vanishing cosmological constant. Therefore, it follows that, within 
string theory, the naturalness principle of 't Hooft \cite{Hooft} can be 
satisfied as the vanishing of the cosmological constant implies that the 
theory has more symmetry, namely $S$-duality. Our work shows however that 
the situation is more complex, as the cosmological constant has to arise from 
a potential term and this should be, as we have been explicitly considering, 
modular invariant, that is invariant under $SL(2,\bf{Z})$ transformations. 
Furthermore, the process of symmetry breaking has also to be taken into 
account. From Figure 1 one sees that the potential has an infinite number of 
minima located at the points ${\bf Im}~S \in {\bf Z}$ and ${\bf Re}~S=0.8$ or 
${\bf Re}~S=1.3$. Indeed, as the components of the $S$ field settle in the 
ground state, $SL(2,\bf{Z})$ is broken down to ${\bf Z}_{\infty}$. After this 
spontaneous symmetry breaking of modular invariance, we still have axion 
fluctuations about the degenerate minima. Recall that $\chi={\bf Im}~S$ is 
the axion field, and one can represent axion oscillations about the several 
minima as governed by the following potential, near all $n$ and near both 
$\langle {\bf Re}~S \rangle$, written as (see Figure 1):

\beq
V(S,S^{+}) = \alpha(\langle {\bf Re}~S \rangle)~[{\bf Im}~S - n]^{2}~,~~~~n 
\in {\bf Z}~,
\label{eq:3.7}
\eeq 

\noindent
where $\alpha(\langle {\bf Re}~S \rangle) = \{\alpha(0.8), \alpha(1.3)\} = 
\{\alpha_{-},\alpha_{+}\} \equiv \alpha(j)$. So, the potential depends on the 
integer $n$ and the choice of factor $\alpha(j)$. We can therefore label 
each vacua as $\vert jn \rangle$, where the potential is $V_{jn}=
\alpha(j)[\chi-n]^2$.

Moreover, this implies that we have a $\theta$-vacua which can be
labeled by the quantum numbers $\vert jn \rangle$. Thus, the ground-state 
wave function is a quantum superposition over all the absolute minima 
of the modular invariant scalar potential:

\beq
\Psi (a,S,S^{+}) = \sum_{j = \pm} \sum_{n \in {\bf Z} } c_{jn} 
\Psi_{jn} (a,S,S^{+})~.
\label{eq:3.8}
\eeq

The wave function for each state $\vert jn \rangle$ can be computed with the
help of the WKB approximation (and in the late Universe approximation), 
that is 

\beq
\Psi^{(+)}(a,S,S^{+}) = N \exp\{i S(a,S,S^{+})/ {\hbar}\}.
\label{eq:3.9}
\eeq

At one-loop level the wave function can be easily obtained:
 
\beq 
\Psi_{jn}^{(+)}(a,S,S^{+})  =  \frac{N(S,S^{+})}{a^{1/2}~
[\langle \hat V_{jn} \rangle a^2 + 6]^{1/4}} 
\exp \left\{-{4~\sqrt{6}~\pi^{2} \over 3~
{\hbar}~\langle \hat V_{jn} \rangle}~
[\langle \hat V_{jn} \rangle a^2 + 6]^{3/2} + {\cal O}({\hbar})\right\}~,
\label{eq:3.10}
\eeq

\noindent
where $N(S,S^{+})$ is a normalization constant that depends on values of 
$S$ and $S^{+}$ at the minimum.

The most striking feature of this wave function is that it is sharply peaked
at $\langle \hat V \rangle \rightarrow 0^{-}$, implying that the most likely 
field configurations for expanding Universes are the ones consistent with this 
condition. Of course, this is reminiscent of the earlier ideas of Baum,
Hawking and Coleman \cite{Baum,Hawking2,Coleman}. In \cite{Coleman} it was 
shown that the inclusion of wormhole type solutions in the Euclidean path 
integral allows for treating the contribution of inequivalent topological
configurations to the wave function of the Universe. After using the 
dilute gas approximation to compute the instanton-wormhole contribution
to the Euclidean effective action it was shown that the parameters of the
effective theory, such as the cosmological constant, the gravitational 
constant, etc., turn into dynamical variables whose values are fixed 
once the wave function is maximized. It was also argued by Coleman that the
inclusion of the wormhole contribution to the Euclidean path integral would
fix the normalization problem that rendered the proposal of 
Baum and Hawking inconsistent. This point however, has been criticized 
on various grounds, specially in what concerns the fact that Euclidean 
quantum gravity is, for a vanishing cosmological constant, unbounded from 
below from which it follows that the theory does not have a stable ground 
state (see {\it e.g.} Ref. \cite{Fischler} for a thorough discussion). 
Despite all these issues, we stress that our analysis  
indicates that $S$-modular invariance at the quantum level implies that 
the cosmological constant problem in ${\cal N}=1$ supergravity strongly 
resembles the strong CP problem. In the latter, the $\theta$ angle is 
adjusted to vanish thanks to the Peccei-Quinn field associated to an 
extra $U(1)_{PQ}$ symmetry. Adjusting mechanisms for the vanishing of the 
cosmological constant inspired on the Peccei-Quinn mechanism have been 
envisaged \cite{Wilczek}, although strong arguments on the lack of 
effectiveness of such mechanisms have been put forward \cite{Weinberg}.  

Thus, we have seen in this section that our analysis of the solutions of
Wheeler-DeWitt equation for a modular invariant ${\cal N} = 1$ supergravity
model in an homogeneous and isotropic spacetime does shed some light into
the issue of the initial conditions for the onset of energy density 
fluctuations generated at the topological inflationary scenario discussed in 
Ref. \cite{Bento3} and also into the problem of the cosmological constant 
and its striking resemblance with the strong CP problem. 

\vspace*{0.5cm}

\section{Interpretation of the Wave Function}

\indent

The behavior of the wave function we have obtained in the previous section 
can be analyzed using the square of the trace of the extrinsic curvature, 
$K^2=K_{ij}K^{ij}$, which allows establishing whether the wave function in 
the semiclassical limit corresponds to a Lorentzian or to an Euclidean 
geometry. Of course, the Wheeler-DeWitt equation is the same whether 
Lorentzian or Euclidean metrics are used to derive it. The extrinsic 
curvature is a measure of the variation of the normal to the hypersurfaces 
of constant time, and is given in general by:

\beq
K_{i j}=N^{-1}\left(-\frac{1}{2}\frac{\partial h_{i j}}
{\partial t}+\nabla_{j}N_{i}\right)~,
\label{eq:4.1}
\eeq

\noindent
where $h_{ij}$ is the $3$-metric and $N_{i}$ are the components of the 
shift-vector. For our geometry (\ref{eq:2.2}) we have

\beq
K_{i j}= - \frac{\dot a}{a} h_{i j}~,
\label{eq:4.2}
\eeq

\noindent
and

\beq
K^2= - \frac{1}{192 \pi^{4} a^{4}} {\partial^2 \over \partial a^{2}}~.
\label{eq:4.3}
\eeq

Hence, in order to interpret the wave function we have to consider the 
following quantity:

\beq
W \Psi(a,S,S^{+}) \equiv \frac{K^2 \Psi(a,S,S^{+})}{\Psi(a,S,S^{+})}~.
\label{eq:4.4}
\eeq

If $W$ is positive it implies the wave function is oscillatory and therefore 
it corresponds to a classical and Lorentzian geometry. If, on the other hand, 
$W$ is negative, then the wave function is ``exponential'' or of tunneling 
type corresponding to a quantum or Euclidean geometry. A rather 
straightforward computation indicates that $W \Psi(a,S,S^{+})$ behaves as 
follows:  

\noindent
{\sl Very Early Universe:}
\beq
W \Psi^{(--)}(a,S,S^{+}) <  0~;
\label{eq:4.5}
\eeq

\noindent
{\sl Early Universe:}
\beq
W \Psi^{(-)}(a,S,S^{+}) <  0~;
\label{eq:4.6}
\eeq

\noindent
{\sl Late Universe:}
\beq
W \Psi^{(+)}(a,S,S^{+}) > 0~.
\label{eq:4.7}
\eeq

\noindent
These results indicate that the transition from the quantum to 
classical regime occurs for $a > 1$ after which the quartic term in the 
scale factor in the minisuperspace potential, 
$U(a,S,S^{+}) \equiv - 96\pi^{4}a^{2}(6 + a^{2}~\hat V(S,S^{+}))$, 
dominates the quadratic one arising from the spatial curvature.
Moreover, in the oscillatory or classical region the wave function can be 
further analyzed using the WKB approximation, $\Psi = C e^{iI}$, where 
$C$ is a slowly varying factor and $I$ a rapidly varying phase. 
The phase $I$ is chosen so that it satisfies the Hamilton-Jacobi equation:

\beq
- \left(\frac{\partial I}{\partial a}\right)^2 + 
\frac{24~(S + S^{+})^2}{a^2} \left(\frac{\partial I}{\partial S~}\right) 
 \left(\frac{\partial I}{\partial S^{+}}\right) + U(a,S,S^{+}) = 0~.
\label{eq:4.8}
\eeq

The meaning of the phase $I$ can be understood acting with the operators 
$\Pi_a$, $\Pi_S$ and $\Pi_{S^{+}}$ on the wave function. Indeed, for instance, 
operating with $\Pi_a$ (the procedure is analogous for $\Pi_S$ and 
$\Pi_{S^{+}}$) yields:

\beq
\Pi_a \Psi = \left[\frac{\partial I}{\partial a} -  
i \frac{\partial \ln C}{\partial a}\right] \Psi~,
\label{eq:4.9}
\eeq
and one realizes that, since in the WKB approximation

\beq
\left\vert \frac{\partial I}{\partial a}\right\vert \gg
\left\vert \frac{\partial}{\partial a}\ln C\right\vert~~,
\label{eq:4.10}
\eeq
then
\beq
\Pi_{a}=\frac{\partial I}{\partial a}~~,~~
\Pi_{S}=\frac{\partial I}{\partial S}~~,~~
\Pi_{S^{+}}=\frac{\partial I}{\partial S^{+}}~~.
\label{eq:4.11}
\eeq

\noindent
Hence the wave function corresponds in this situation to a three-parameter
subset of solutions of (\ref{eq:4.8}) and can be interpreted as a 
boundary condition for the classical solutions.

\vspace*{0.5cm}

\section{Discussion and Conclusions}

\indent

In this paper we have obtained solutions of the Wheeler-DeWitt equation 
from the minisuperspace model resulting from imposing $S$-modular 
invariance in the bosonic sector of ${\cal N} = 1$ supergravity, 
assuming a closed Friedmann-Robertson-Walker 
spacetime. In section 2 we have introduced the specificities of the model 
that were associated with the modular invariance, obtained the Wheeler-DeWitt 
equation of our problem and outlined our strategy in searching for its 
solutions. We have implemented the no-boundary Hartle-Hawking proposal in 
section 3 and addressed the issues of initial field configurations at the 
very early Universe and of the cosmological constant. The former discussion 
is, of course, relevant when studying the onset of inflation and the very 
early Universe conditions that have given rise to the energy density 
fluctuations and gravitational waves generated by inflation. We have used the 
scalar-factor duality to obtain the wave function for this case and have shown 
that, for an expanding Universe, the most likely configuration for the $S$ 
field was the one where it started sitting at the top of the modular 
invariant ${\cal N} = 1$ supergravity potential. This is consistent with 
assumptions assumed in Ref. \cite{Bento3} in the context of a topological 
inflationary model built in an $S$ and $T$ dual ${\cal N} = 1$ supergravity 
model. In order to address the cosmological constant problem we have 
calculated the wave function of the Universe in the limit of very large 
scale-factor and we have shown that the wave function is sharply peaked, 
after spontaneous symmetry breaking of the $SL(2,\bf{Z})$ modular invariance 
down to ${\bf Z}_{\infty}$, at a vanishing potential from below. This feature 
is common to known solutions for the cosmological constant problem 
\cite{Baum,Hawking2,Coleman}, but with the novelty that, given the 
ground-state structure of dual ${\cal N} = 1$ supergravity, there is actually 
a $\theta$-vacuum which suggests that the cosmological constant problem has, 
at the quantum level, a great resemblance with the strong CP problem.

Finally,  we have identified, in section 4, the regimes where the 
wave function is Lorentzian or Euclidean applying the operator 
$K^2=K_{ij}K^{ij}$, where 
$K_{ij}$ is the extrinsic curvature of spacetime manifold, on the wave 
function. We have further interpreted the classical Lorentzian regime 
using the WKB approximation and writing the Hamilton-Jacobi equation that the
relevant phase must satisfy. We have found that the Universe evolved from a 
quantum to a classical 
regime after the scale factor quartic term in the minisuperspace potential 
dominated the quadratic term. We have also shown that our analysis 
succeeded in both 
establishing a quantum mechanical validation of the classical treatment 
of Refs. \cite{Bento2,Bento3} concerning topological inflation, 
as well as hinting 
some directions in a possible mechanism for explaining 
the vanishing of the cosmological constant in the late Universe.

\vspace{1cm}

{\large\bf Acknowledgments} 

\noindent
One of us (R.S.) is supported in part by funds provided by the U.S. 
Department of Energy (D.O.E.) under cooperative research agreement 
$\sharp$DE-FC02-94ER40818, in part by Funda\c c\~ao Calouste Gulbenkian 
(Portugal), and in part by the Funda\c c\~ao Luso-Americana para o 
Desenvolvimento grant 50/98 (Portugal).

\newpage

\newcommand\figinsert[4]
{\begin{figure}[htb]
\newcommand\figsize{#3}
\epsfysize\figsize
\vskip -.5cm
\centerline{\epsffile{#4}}
\vskip -.0cm
\caption{\leftskip 1pc \rightskip 1pc
\baselineskip=10pt plus 2pt minus 0pt {{#2}}}
\label{fig #1}
\vskip -.3cm\end{figure}}


\figinsert{Fig1}                    
{The scalar $S$-dual potential $V(S,S^{+})$, Eq. (5), as a function of 
$({\bf Re}~S,{\bf Im}~S)$.}         
{4.0truein}{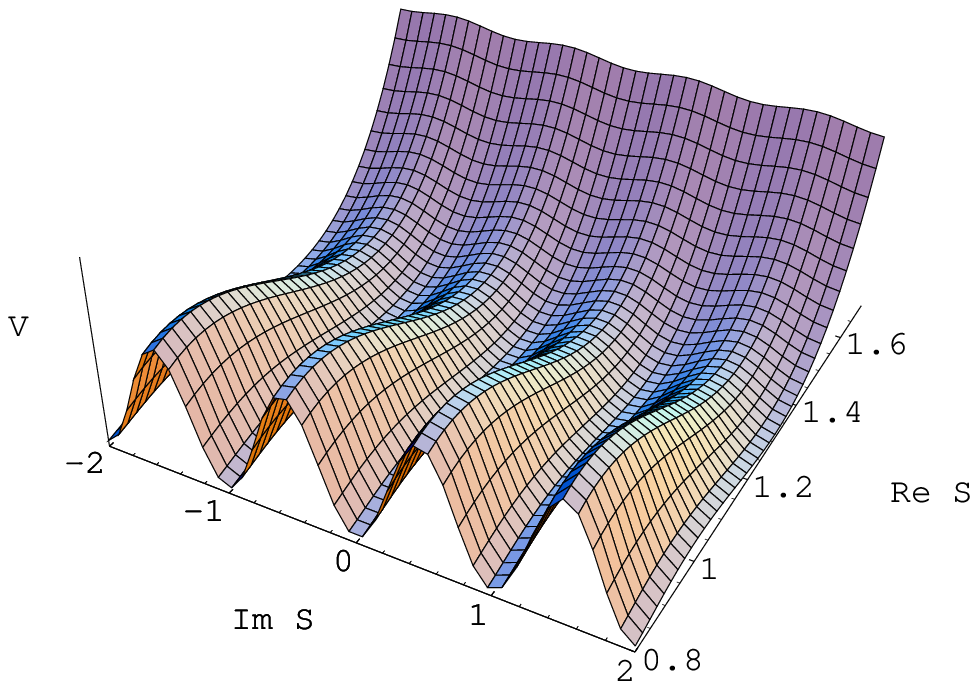}             

\end{document}